\documentclass[10pt,twocolumn,oneside,journal]{IEEEtran}
\usepackage{cite}
\usepackage{graphicx}
\usepackage{amsmath}
\usepackage{times}
\usepackage{latexsym}
\usepackage{graphicx}
\usepackage{bm}
\usepackage{amssymb}
\usepackage[center]{caption2}
\usepackage{stfloats}
\usepackage{cases}
\usepackage{array}
\usepackage{setspace}
\usepackage{fancyhdr}
\usepackage{fancyhdr}
\usepackage{algorithm}
\usepackage{algpseudocode}
\usepackage{textcomp}
\usepackage{wasysym}
\usepackage{url}
\usepackage{xcolor}
\usepackage{stfloats}
\usepackage{amsmath,epsfig,amssymb,bm,dsfont}
\usepackage{lipsum}

\graphicspath{{figures/}}

\allowdisplaybreaks[4]

\newcaptionstyle{mystyle1}{%
  \centering TABLE  \captiontext \par}
\captionstyle{mystyle1}
\newcaptionstyle{mystyle2}{%
  \captionlabel $.$ \: \captiontext \par}
\captionstyle{mystyle2}
\newcaptionstyle{mystyle3}{%
  \captionlabel $.$ \: \captiontext \par}
\captionstyle{mystyle3}

\renewcommand{\QED}{\QEDopen}
\begin{document}
\title{Cooperative Caching in Fog Radio Access Networks: A Graph-based Approach}


\author{Yanxiang~Jiang,~\IEEEmembership{Senior~Member,~IEEE},
Xiaoting~Cui, Mehdi~Bennis,~\IEEEmembership{Senior~Member,~IEEE}, and~Fu-Chun~Zheng,~\IEEEmembership{Senior~Member,~IEEE} 
\thanks{This work was supported in part by
the Natural Science Foundation of China under Grant 61521061,
the Natural Science Foundation of Jiangsu Province under grant
BK20181264,
the Research Fund of the State Key Laboratory of
Integrated Services Networks (Xidian University) under grant ISN19-10,
the Research Fund of the Key Laboratory of Wireless Sensor Network $\&$ Communication (Shanghai Institute of Microsystem and Information Technology, Chinese Academy of Sciences) under grant 2017002,
the National Basic Research Program of China
(973 Program) under grant 2012CB316004,
and the U.K. Engineering and Physical Sciences Research Council under Grant EP/K040685/2.}
\thanks{Y. Jiang is with the National Mobile Communications Research Laboratory, Southeast University, Nanjing 210096, China,
the State Key Laboratory of Integrated Services Networks,  Xidian University, Xi'an 710071, China,
and the Key Laboratory of Wireless Sensor Network $\&$ Communication, Shanghai Institute of Microsystem and Information Technology,
Chinese Academy of Sciences, 865 Changning Road,
Shanghai 200050, China (e-mail: yxjiang@seu.edu.cn).}
\thanks{X. Cui is with the National Mobile Communications Research Laboratory, Southeast University, Nanjing 210096, China. 
}
\thanks{M. Bennis is with the Centre for Wireless Communications, University of Oulu, Oulu 90014, Finland (e-mail: mehdi.bennis@oulu.fi).}
\thanks{F. Zheng is with the School of Electronic and Information Engineering, Harbin Institute of Technology, Shenzhen 518055,  China,
and the National Mobile Communications Research Laboratory, Southeast University, Nanjing 210096, China (e-mail: fzheng@ieee.org).}
}

%

\maketitle

\begin{abstract}
In this paper, cooperative caching is investigated in fog radio access networks (F-RAN). To maximize the offloaded traffic,  cooperative caching optimization problem is formulated. By analyzing the relationship between clustering and cooperation and utilizing the solutions of the knapsack problems, the above challenging optimization problem is transformed into a clustering subproblem and a content placement subproblem. To further reduce complexity, we propose an effective graph-based approach to solve the two subproblems. In the graph-based clustering approach, a node graph and a weighted graph are constructed. By setting the weights of the vertices of the weighted graph to be the incremental offloaded traffics of their corresponding complete subgraphs, the objective cluster sets can be readily obtained by using an effective greedy algorithm to search for the max-weight independent subset. In the graph-based content placement approach, a redundancy graph is constructed by removing the edges in the complete subgraphs of the node graph corresponding to the obtained cluster sets. Furthermore, we enhance the caching decisions to ensure each duplicate file is cached only once. Compared with traditional approximate solutions, our proposed graph-based approach has lower complexity. Simulation results show remarkable improvements in terms of offloaded traffic by using our proposed approach.
\end{abstract}


\section{Introduction}
With the continuous and rapid proliferation of various intelligent devices and advanced mobile application services, wireless networks have been suffering an unprecedented data traffic pressure in recent years.
Ever-increasing mobile data traffic brings tremendous load on capacity-limited fronthaul links, especially at peak traffic moments.
As a promising architecture, fog radio access networks (F-RAN) can effectively offload the traffic  in fronthaul links by placing popular contents at fog access points (F-APs) which are equipped with limited caching resources \cite{Peng}.
Due to storage constraint and fluctuant spatio-temporal traffic demands, cooperative caching is an effective way to increase the offloaded traffic.

Recently, there have been a lot of  works on cooperative caching.
In \cite{WJ}, a cooperative caching and delivery policy was proposed to minimize the latency, where each base station (BS) and user equipments (UEs) cached files according to the request probability independently.
However, the caching decision of one BS was influenced by that of the neighboring cooperative BSs, and different BSs should cache diverse files in a cooperative manner \cite{XLi, SZ}.
In \cite{TQ,KPoularakis,Sun}, the cooperative content placement strategy for the given cache nodes cluster was studied.
In \cite{TQ}, a cooperative content placement strategy was proposed to maximize the service probability, where the storage space of each BS in the given cluster was divided into a proportion for caching the same files and a rest proportion for caching different files.
In \cite{KPoularakis}, a cooperative caching algorithm for multiple operators was proposed to maximize the delay savings, where all the cache nodes in the given cluster firstly cached the globally popular files together and then cached the locally popular files independently.
In \cite{Sun}, a cooperative  content placement method was proposed to minimize the latency for multi-cell cooperative networks, where a heuristic greedy algorithm with limited performance guarantee was developed.
In \cite{KShanmugam,Zheng,ElBamby}, the cooperative content placement strategy for unknown cache nodes cluster was studied.
In \cite{KShanmugam}, the uncoded and coded cooperative content assignment strategies were proposed to minimize the expected downloading time, where the connectivity graph between UEs and BSs was used to reflect cooperation relationship among neighboring BSs.
By optimizing relay clustering and content placement in a joint manner, a cooperative caching strategy was developed to minimize the outage probability in \cite{Zheng},  where identical files were  cached among the relays in each cluster for simplicity.
Based on the similarities among users requesting similar contents, a user clustering and cooperative caching algorithm to improve the cache hit rate was proposed in \cite{ElBamby}.

However, the prior works on cooperative content placement tend to exploit the global content popularity rather than  the local content popularity, which might not even replicate the  global content popularity. The local content popularity indeed reflects user interest at the coverage of each cache node and might be different from each other \cite{Atan, add10}.
It was investigated in \cite{STamoor} and \cite{Liu} that the cooperative content placement algorithms based on the local content popularity could obtain lower delay or higher cache hit rate than that based on the global content popularity.

Motivated by the aforementioned discussions, the main contributions of this paper are summarized below.
\begin{itemize}
\item{We propose a new idea for solving the challenging cooperative caching optimization problem based on the local content popularity. Analyzing the relationship between clustering and cooperation and utilizing the solutions of the knapsack problems, we transform the cooperative caching optimization problem into a clustering subproblem and a content placement subproblem.}
\item{We propose a graph-based clustering approach. Constructing a node graph and a weighted graph, we transform the clustering subproblem into an equivalent 0-1 integer programming problem. Furthermore, we propose an effective greedy algorithm to search for the objective cluster sets.}
\item{We propose a graph-based content placement approach. Constructing a redundancy graph based on the obtained cluster sets, we determine the duplicate files that will indeed cause cache redundancy at each edge and further enhance the caching decisions for each file. Correspondingly, all the possible cache redundancy can be  eliminated by caching each duplicate popular file only once.}
\end{itemize}

The rest of this paper is organized as follows.
In Section II, the system model and problem formulation are briefly described.
In Section III, the problem transformation is presented.
The proposed graph-based cooperative caching scheme including clustering and content placement is presented in Section IV. Simulation results are shown in Section V.
Final conclusions are drawn in Section VI.

\section{System Model and Problem Formulation}
\begin{figure}[!t]
\centering 
\includegraphics[width=0.43\textwidth]{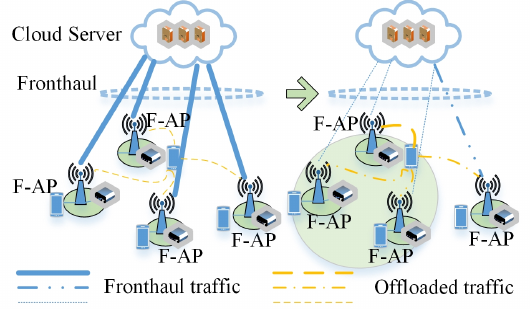}
\caption{{Illustration of the cooperative caching scenario in F-RAN.}}
\label{System}
\end{figure}

Consider a cooperative caching scenario in F-RAN as illustrated in Fig. \ref{System}, which consists of a cloud server, ${M}$ F-APs, and a certain number of users. {The cloud  server can be accessed by the F-APs via fronthaul links.
Let  ${\cal M} = \left\{ {1,2,\cdots,m,\cdots,M} \right\}$ denote the F-AP set.}
Assume that neighboring F-APs can share files and cooperate with each other \cite{JL}. Whether two F-APs can cooperate or not depends on how well they satisfy some certain rules.
Let ${{\cal S}_m}$ denote the set of all the cooperators of F-AP $m$.
Without loss of generality, assume that all the  files have the same size of $L$ bits, each F-AP has the same storage space and can store up to $K $ files from the content library ${\cal F} = \left\{ {1,2,\cdots,f,\cdots,F} \right\}$ located in the cloud server.
Let ${p_{m{f}}}$ denote the request probability of file ${f}$ at F-AP $m$ (referred to as the local content popularity).
Assume that the request probability at each F-AP is stationary during the given time period.
Let ${\lambda _m}$ denote the aggregate request arrival rate at F-AP $m$,
and ${w_m} = \lambda _m / {\sum\nolimits_{m' \in \mathcal M} {{\lambda _{m'}}} }$ denote
the ratio of the traffic load at F-AP $m$ to the sum load of the $M$ F-APs.

Let ${x_{mf}} \in \left\{ {1, 0} \right\}$ denote the caching decision of file $f$ at F-AP $m$, where $x_{mf}=1$ if file $f$ is cached at F-AP $m$ and $x_{mf}=0$ otherwise. Let $x_{mf}^{\text l}\in \left\{ {1, 0} \right\}$ denote the local state of  file $f$ at F-AP $m$  and its cooperators, where $x_{mf}^{\text l}=1$ if file $f$ is successfully cached locally; $x_{mf}^{\text l}=0$ if file $f$ is not cached locally and must be fetched  from the cloud  server. Then, $x_{mf}^{\text l}$ can be expressed as follows:
\begin{equation}\label{Sys_xmf^a}
x_{mf}^{\text l} = {x_{mf}} + \left( {1 - {x_{mf}}} \right)\left[ {1 -  \prod\nolimits_{m' \in {{\cal S}_m}} {\left( {1 - {x_{m'f}}} \right)} } \right].
\end{equation}

\newcounter{TempEqCnt}
\setcounter{TempEqCnt}{\value{equation}}
\setcounter{equation}{13}
\begin{figure*}[!b]
\hrulefill
\begin{multline}\label{Sys_xmf^a2}
x_{mf}^{\text l}  = {x_{mf}} + \left( {1 - {x_{mf}}} \right)\left[ {1 -  \prod\nolimits_{m' \in {{\cal S}_m^1}} {\left( {1 - {x_{m'f}}} \right)} } \right] + \\
\left( {1 - {x_{mf}}} \right) \prod\nolimits_{m' \in {{\cal S}_m^1}} {\left( {1 - {x_{m'f}}} \right)}\left[ {1 -  \prod\nolimits_{m' \in {{\cal S}_m^2 \cup {\cal S}_m^3}} {\left( {1 - {x_{m'f}}} \right)} } \right].
\end{multline}
\end{figure*}
\setcounter{equation}{\value{TempEqCnt}}

Once the requested file is cached locally, the traffic in the fronthaul links can be offloaded. Let $T$ denote the   offloaded traffic for all the considered $M$ F-APs. Then, it can be expressed as follows:
\begin{align}
T &= \sum\nolimits_{m \in \mathcal M} {\sum\nolimits_{f  \in \mathcal F} {{\lambda _m}{p_{mf}}{x_{mf}^{\text l}}L} }. \label{T-new}
\end{align}
Note that the   offloaded traffic increases with the number of locally cached files, and decreases with duplicate cached files at the requested F-APs and their cooperators.
The caching decisions should be determined cooperatively by the neighboring F-APs for a larger number of unduplicated cached files.

To maximize the   offloaded traffic, the cooperators should be neighboring F-APs with closer distance and greater load difference.
The selected F-APs can efficiently offload traffic among each other and are more likely to cooperative with each other \cite{Ashraf}. Let $d_{{m}}$ denote the geographical coordinate of F-AP $m$ in the Euclidean space, $D_{{m}{m'}} = \left\| {{d_{{m}}} - {d_{{m'}}}} \right\|_2$ denote the distance between F-AP $m$ and F-AP $m'$, and $L_{{m}{m'}} = \left\| {{\lambda _{{m}}} - {\lambda _{{m'}}}} \right\|_2$ denote the load difference between F-AP $m$ and F-AP $m'$.
Then, the cooperative caching optimization problem can be formulated as follows:
\begin{align}\label{P-0}
&{\mathop {\text{max} }\limits_{{x_{mf}}}}  { {\; T } } \\
{\text{s.t.}}\
&{D_{{m}{m '}} \le {\gamma ^\text{d}},\ \forall m \in {\cal M}, \forall m' \in {{\cal S}_{m}},} \tag{\ref{P-0}a}\nonumber\\
&{L_{{m}{m'}} \ge {\gamma ^\text{l}},\ \forall m \in {\cal M}, \forall m' \in {{\cal S}_{m}}, } \tag{\ref{P-0}b} \nonumber \\
&{{x_{mf}} \in \left\{ {1, 0} \right\},\ \forall m \in {\cal M},\forall f \in {\cal F},}\tag{\ref{P-0}c}\nonumber \\
&{\sum\nolimits_{f \in \mathcal F} {{x_{mf}}}  \le K, \ \forall m \in {\cal M},}\tag{\ref{P-0}d}\nonumber
\end{align}
where ${{\gamma ^\text{d}}}$  and ${{\gamma ^\text{l}}}$ denote the distance threshold and the load threshold, respectively.

The objective of this paper is to find the optimal caching decisions $\left\{ {{x_{mf}}\left| m \in {\cal M}, f\in {\cal F} \right.} \right\}$ by maximizing the offloaded traffic using cooperative caching in F-RAN.

\section{Problem Transformation}
The optimization problem in (\ref{P-0}) is a 0-1 integer programming problem, which is NP-hard \cite{WJ, KPoularakis}. A dynamic programming approach is generally required for obtaining a global optimal solution  \cite{RWang}. However, such an approach has an exponential complexity with respect to (w.r.t.) the number of F-APs and the size of the content library, and it is computationally impracticable even for a small size network. In the previous works in \cite{KShanmugam, RWang}, by reformulating the original problem into a matroid constrained monotone submodular optimization problem, the approximate solutions with limited performance can be obtained.
However, by using the above approach, it incurs a long running time to evaluate the marginal value of the objective function.

In fact, by utilizing the relationship between clustering and cooperation,
the cooperators of an F-AP can be divided into intra-cluster cooperators, inter-cluster cooperators, and nonclustered cooperators. Correspondingly, the objective function of the cooperative caching optimization problem in (\ref{P-0}) can be decomposed into three items.
All the three items indicate that the offloaded traffic is affected by the clustering strategy. In addition, the first item indicates that the offloaded traffic is also affected by the cached files at the requested F-APs and their intra-cluster cooperators. The second item indicates that the offloaded traffic is also affected by the cached files at the nonclustered cooperators and the inter-cluster cooperators of the requested F-APs. The third item indicates that the offloaded traffic is also affected by the duplicate cached files between the requested F-APs (or their intra-cluster cooperators) and their inter-cluster cooperators. In summary, all the three items indicate that the solution of the original optimization problem requires to determine clusters and content placement.
Therefore, in this paper, we propose to transform the challenging cooperative caching optimization problem into a clustering  subproblem and a content placement subproblem.



%
\setcounter{TempEqCnt}{\value{equation}}
\setcounter{equation}{16}
\begin{figure*}[!b]
\hrulefill
\begin{multline}\label{T-new2}
T = \sum\nolimits_{n \in \mathcal N}\sum\nolimits_{m \in {\cal M}_n^{\text c}}   { {\sum\nolimits_{f \in \mathcal F} {{\lambda _m}{p_{mf}} \left\{ {{x_{nf}+\left( {1 - {x_{nf}}} \right)  \left[ {1 - \prod\nolimits_{m' \in {\cal S}_m^2 \cup {\cal S}_m^3} {{\left( {1 - {x_{m'f}}} \right)} } } \right]}} \right\} L} } }
  + \\
  \sum\nolimits_{m \in {{\cal M}^{\text n}}} {\sum\nolimits_{f  \in \mathcal F} {{\lambda _m}{p_{mf}} \left\{ {x_{mf}+\left( {1 - {x_{mf}}} \right)
 \left[ {1 - \prod\nolimits_{m' \in {\cal S}_m^2 } {\left( {1 - {x_{m'f}}} \right)} } \right]} \right\} L} }. 
\end{multline}
\end{figure*}

\setcounter{equation}{\value{TempEqCnt}}
\setcounter{TempEqCnt}{\value{equation}}
\setcounter{equation}{19}
\begin{figure*}[!b]
\vspace*{-5pt}
\begin{multline} \label{T_n}
{T^{\text n}}=\sum\nolimits_{n \in \mathcal N} {\sum\nolimits_{m \in {\cal M}_n^\text{c}} {\sum\nolimits_{f \in \mathcal F} {{\lambda _m}{p_{mf}}\left[ {1 - \prod\nolimits_{m' \in {\cal S}_m^2 \cup {\cal S}_m^3} { {\left( {1 - {x_{m'f}}} \right)} } } \right]L} } } + \\
\sum\nolimits_{m \in {{\cal M}^\text{n}}} {\sum\nolimits_{f \in \mathcal F} {{\lambda _m}{p_{mf}}\left[ {1 - \prod\nolimits_{m' \in {\cal S}_m^2 } {\left( {1 - {x_{m'f}}} \right)} } \right]L} }.
\end{multline}
\begin{multline} \label{T_s2s3}
{T^\text{d}}=\sum\nolimits_{n \in \mathcal N} \sum\nolimits_{m \in {\cal M}_n^\text{c}} { {\sum\nolimits_{f \in \mathcal F} {{\lambda _m}{p_{mf}}{x_{nf}}\left[ {1 - \prod\nolimits_{m' \in {\cal S}_m^2 \cup {\cal S}_m^3} {{\left( {1 - {x_{m'f}}} \right)} } } \right]L}}}
+\\
\sum\nolimits_{m \in {{\cal M}^\text{n}}} {\sum\nolimits_{f \in \mathcal F} {{\lambda _m}{p_{mf}}{x_{mf}}\left[ {1 - \prod\nolimits_{m' \in {\cal S}_m^2 } {\left( {1 - {x_{m'f}}} \right)} } \right]L} }.
\end{multline}
\end{figure*}
\setcounter{equation}{\value{TempEqCnt}}

\vspace*{-5pt}

\subsection{Clustering and Cooperation}
Cooperative F-APs can form a cluster to make the storage space in a cluster be seen as an entirety \cite{Chen}. Correspondingly, clustering can increase the content diversity.
Any two F-APs in a cluster can cooperate with each other.
However, two F-APs that can cooperate may not necessarily be members of a cluster.

Assume that the considered $M$ F-APs can constitute $N$ disjoint clustered sets denoted by ${{\cal M}^\text{c}_n}$ for $n \in \mathcal N = \{1,2, \cdots, N\}$ and one nonclustered set denoted by ${{\cal M}^\text{n}}$, and the set size of ${{\cal M}^\text{c}_n}$ is denoted by  $S_n$.
{Disjoint clustering makes one F-AP  only be a member of one cluster, which ensures exclusive and sufficient usage of its storage space to all the users in the cluster.}
Correspondingly, the following relationship can be readily established:
\begin{equation}
{\cal M} =  \left( { \cup _{n \in \mathcal N}}{{{\cal M}^\text{c}_n}} \right) \cup {\cal M}^\text{n}, \label{M}
\end{equation}
\begin{equation}\label{eq5}
{{\cal M}_n^\text{c}} \cap {{\cal M}_{n'}^\text{c}} = \varnothing ,\   \forall  n, n' \in \mathcal N, n \ne {n'}.
\end{equation}
Without loss of generality, let ${\cal S}_m^1$, ${\cal S}_m^2$, and ${\cal S}_m^3$ denote the set of intra-cluster cooperators, inter-cluster cooperators, and nonclustered cooperators of F-AP $m$, respectively.
Define
\begin{equation}
{{\cal S}_m } = {\cal S}_m^1 \cup {\cal S}_m^2 \cup {\cal S}_m^3. \label{Sm}
\end{equation}
Then, the following relationship can be readily established:
\begin{align}
{\cal S}_m^i \cap {\cal S}_m^j &= \varnothing,\  \forall i, j \in \left\{1,2,3\right\}, i \ne j, \\
m \cup {\cal S}_m^1 &= {\cal M}_n^{\text c},\ m \in {\cal M}_n^{\text c}, \label{m-Sm1} \\
{\cal S}_m^1 & = {\cal S}_m^3 = \varnothing,\ m \in {\cal M}^{\text n}. \label{m^n-Sm1}
\end{align}

Let ${p_{n{f}}}$ denote the request probability of file ${f}$ in cluster ${n}$. Then, according to \cite{Li}, we have:
\begin{equation}
{p_{n{f}}} = \sum\nolimits_{m \in {{\cal M}^\text{c}_n}} {{p_{m{f}}}\frac{w_m}{\sum\nolimits_{m' \in {{\cal M}_n^\text{c}}} {{w _{m'}}} }}.
\end{equation}
Assume that cluster ${n}$ can cache $K_n = S_n  K$ different files.
Generally, ${S_n}K \ll F$.
Let $x_{nf}\in \left\{ {1, 0} \right\}$ denote the caching decision of file $f$ in cluster $n$, where $x_{nf}=1$ if file $f$ is cached at any F-AP in cluster $n$ and $x_{nf}=0$ otherwise.
Then, we have:
\begin{align}
x_{nf} &={1 - \prod\nolimits_{m \in {\cal M}_n^c} {\left( {1 - {x_{mf}}} \right)} }, \label{Sys_xnf-1} \\
&=1-\left( {1 -  {x_{mf}}} \right)\prod\nolimits_{m' \in {{\cal S}_m^1}} {\left( {1 - {x_{m'f}}} \right)},   {m \in {\cal M}_n^c}, \label{Sys_xnf-2}
\end{align}
\begin{equation} \label{x_nfConstraint}
\sum\nolimits_{f \in \mathcal F} {{x_{nf}}}  \le {K_n},\  n \in \mathcal N.
\end{equation}

\subsection{Objective Function Decomposition}

Substituting (\ref{Sm}) into (\ref{Sys_xmf^a}), the local state $x_{mf}^{\text l}$ of the requested file $f$ at F-AP $m \in {\cal M}$ and its cooperators can be expressed in an equivalent form in (\ref{Sys_xmf^a2}) as shown at the bottom of this page.
When $m \in {{\cal M}^{\text n}}$, according to (\ref{m^n-Sm1}) and (\ref{Sys_xmf^a2}), $x_{mf}^{\text l}$  can be further expressed as follows:
\setcounter{equation}{14}
\begin{multline}\label{Sys_xmf^l-n}
x_{mf}^{\text l} = {x_{mf}} + \left( {1 - {x_{mf}}} \right)\left[ {1 - \prod\nolimits_{m' \in {{\cal S}_m^2 }} {\left( {1 - {x_{m'f}}} \right)} } \right],  \\
m \in {{\cal M}^{\text n}}.
\end{multline}
{When $m \in {{\cal M}_n^{\text c}}$, according to (\ref{Sys_xnf-2}) and (\ref{Sys_xmf^a2}), $x_{mf}^{\text l}$ can be further expressed as follows:}
\begin{multline}\label{Sys_xmf^l-c}
x_{mf}^{\text l} = {x_{nf}} + \left( {1 - {x_{nf}}} \right)\left[ {1 - \prod\nolimits_{m' \in {{\cal S}_m^2 \cup {\cal S}_m^3}} {\left( {1 - {x_{m'f}}} \right)} } \right],  \\
m \in {{\cal M}_n^{\text c}},  n \in \mathcal N.
\end{multline}

Substitute (\ref{M}), (\ref{Sys_xmf^l-n}), and  (\ref{Sys_xmf^l-c}) into (\ref{T-new}). Then, the objective function of the original optimization problem in \eqref{P-0} can be expressed in an equivalent form in (\ref{T-new2})  as shown at the bottom of this page.
{For all the considered $M$ F-APs, let ${T^\text{c}}$ denote the offloaded traffic   through fetching files that are cached at the requested F-APs and their intra-cluster cooperators,
${T^{\text n}}$ denote the offloaded traffic  through fetching files that are cached at  {the nonclustered cooperators and the inter-cluster cooperators of the requested F-APs},
and ${T^{\text d}}$  denote the offloaded traffic through  {fetching duplicate files that are cached between the requested F-APs (or their intra-cluster cooperators) and their inter-cluster cooperators}, respectively.}
Then, (\ref{T-new2}) can be decomposed into three items as follows:
\setcounter{equation}{17}
\begin{equation} \label{T-twosubterm}
T =  {T^{\text c}} + {T^{\text n}}-{T^{\text d}},
\end{equation}
where
\begin{multline} \label{T_s1}
{T^\text{c}} =\sum\nolimits_{n \in \mathcal N} \sum\nolimits_{m \in {\cal M}_n^{\text c}} { {\sum\nolimits_{f \in \mathcal F} {{\lambda _m}{p_{mf}}{x_{nf}}L} } }  \\
 +\sum\nolimits_{m \in {{\cal M}^{\text n}}} {\sum\nolimits_{f \in \mathcal F} {{\lambda _m}{p_{mf}}{x_{mf}}L} },
\end{multline}
and ${T^\text{n}}$ and ${T^\text{d}}$ are expressed in (\ref{T_n}) and (\ref{T_s2s3}), respectively, as shown at the bottom of this page.

%

It can be readily seen from \eqref{T_s1} that $T^{\text c}$ can be maximized if the cluster sets are determined, the most popular $K_n$ files in each cluster and the most popular $K$ files at each nonclustered F-AP are cached, respectively.
It can be readily seen from \eqref{T_n} that $T^{\text n}$ can be maximized if the cluster sets are determined, the most popular $K$ files at the inter-cluster cooperators and the nonclustered cooperators of a clustered F-AP are cached at the clustered F-AP, and the most popular $K$ files at the inter-cluster cooperators of a nonclustered F-AP are cached at the nonclustered F-AP.
It can be readily seen from \eqref{T_s2s3} that $T^{\text d}$ can be minimized if the cluster sets are determined, different files are cached between a clustered F-AP and its inter-cluster cooperators (or its nonclustered cooperators), different files are cached between a nonclustered F-AP and its inter-cluster cooperators.

According to the above presentation, firstly, if a clustered F-AP does not have inter-cluster cooperators and nonclustered cooperators, or a nonclustered F-AP does not have inter-cluster cooperators, the cache files at this F-AP cannot be determined through maximizing $T^{\text n}$ whereas they must be determined through maximizing $T^{\text c}$.
Secondly, for the problem of maximizing $T^{\text n}$, the number of most popular files at the inter-cluster cooperators (or the nonclustered cooperators) of a clustered F-AP that should be cached at the clustered F-AP,
and the number of popular files at the inter-cluster cooperators of a nonclustered F-AP that should be cached at the nonclustered F-AP cannot be determined.
It is hardly possible to solve the problem of maximizing $T^{\text n}$.
Thirdly, for the clustered F-APs which have inter-cluster cooperators or nonclustered cooperators, and the non-clustered F-APs which have inter-cluster cooperators, both maximizing  $T^{\text c}$ and maximizing $T^{\text n}$ 
require them to cache the most popular files.
The difference between maximizing  $T^{\text c}$ and maximizing $T^{\text n}$ lies in the caching locations of these files between each pair of a clustered F-AP and its inter-cluster cooperator (or nonclustered cooperator),
and between each pair of a nonclustered F-AP and its inter-cluster cooperator.
There exists an exchange relationship between the caching locations of the above F-AP pairs.
Finally, 
once the popular files
at the clustered and nonclustered F-APs are determined, the duplicate caches files between a clustered F-AP and its inter-cluster cooperators (or noncluster cooperators), and the duplicate caches files  between a nonclustered F-AP and its inter-cluster cooperators can be determined.
By reducing the number of duplicate cached files, caching a duplicate popular file at one F-AP and replacing it by a new popular file at the other F-AP, $T^{\text d}$ can then be minimized.
Based on the above analysis, we 
propose to solve the cooperation caching optimization problem through firstly maximizing $T^{\text c}$ and further minimizing $T^{\text d}$.

\begin{table*}[!t]
\centering
\caption{Summary of major notations}
\begin{tabular}{|c|m{13cm}|}
\hline
\hline
{$M$, $\cal M$, $m$, ${\cal S}_m$, ${\cal S}_m^1$, ${\cal S}_m^2$, ${\cal S}_m^3$}  & {Number of the considered F-APs, set of the $M$ F-APs, index of F-AP, set of all the cooperators of F-AP $m$, set of intra-cluster cooperators of F-AP $m$, set of inter-cluster cooperators of F-AP $m$, set of nonclustered cooperators of F-AP $m$} \\
\hline
$n$, ${{\cal M}^\text{c}_n}$, ${\cal M}^\text n$, $S_n$ &Index of cluster, set of F-APs in cluster $n$,   set of nonclustered F-APs, set size of ${{\cal M}^\text{c}_n}$ \\
\hline
$f$, $\cal F$, $F$ &Index of file, content library, library size \\
\hline
$K$, $K_n$ & Storage size of each F-AP, storage size of cluster $n$\\
\hline
${\lambda _m}$, $w_m$ & Aggregate request arrival rate at F-AP $m$,  ratio of the traffic load at F-AP $m$ to the sum load of the $M$ F-APs\\
\hline
$p_{mf}$, $p_{nf}$,  $p_{{mf}}^\text{o}$, $p_{{nf}}^\text{o}$
& Request probability of file $f$ at F-AP $m$, request probability of file $f$ in cluster $n$,  request probability of the $f$th most popular file at F-AP $m$, request probability of the $f$th most popular file in cluster $n$\\
\hline
$x_{mf}$, $x_{nf}$, $x_{mf}^\text l$ &Caching {decision} of file $f$ at F-AP $m$, caching {decision} of file $f$  in cluster $n$, local state of file $f$ at F-AP $m$ and its cooperators \\
\hline
{$T$, $T^\text c$, $T^\text n$, $T^\text d$} & {Whole offloaded traffic for all the $M$ F-APs, offloaded traffic for all the $M$ F-APs  through fetching files that are cached
at the requested F-APs and their intra-cluster cooperators, offloaded traffic for all the $M$ F-APs through fetching files that
are cached at the nonclustered cooperators and the inter-
cluster cooperators of the requested F-APs, offloaded traffic for all the $M$ F-APs through fetching duplicate files that are
cached between the requested F-APs (or their intra-cluster
cooperators) and their inter-cluster cooperators} \\
\hline
{$T_m$, $T^{\text i}$} & {Offloaded traffic at F-AP m through
fetching files that are cached in its own storage space, incremental offloaded traffic of the $N$ clusters}\\
\hline
$d_m$, $D_{mm'}$,  $L_{mm'}$, ${\gamma ^{\text d}}$, ${\gamma ^{\text l}}$ & Geographical coordinate of F-AP $m$ in the Euclidean space, distance between F-AP $m$ and F-AP $m'$, load difference between F-AP $m$ and F-AP $m'$, distance threshold,  load threshold \\
\hline
\hline
\end{tabular}
\end{table*}

\subsection{Optimization Problem Reformulation}

From (\ref{T_s1}), we can see that $T^\text c$ is affected by the clustering strategy and the caching decisions $\left\{ {{x_{nf}},{x_{mf}}\left| f\in {\cal F},  n \in \mathcal N, m \in {\cal M}^{\text n} \right.} \right\}$. If the cluster sets are determined, $T^\text c$ can be maximized through solving the $N+\left| {\cal M}^\text n \right|$ independent knapsack problems for each $ n \in \mathcal N$ and each $ m \in {\cal M}^\text n$ \cite{WJ}. {Sort $p_{mf}$ and $p_{nf}$ in descending order. Let $p_{{mf}}^\text{o}$ and $p_{{nf}}^\text{o}$ denote the request probability of the $f$th most popular file at F-AP $m$ and in cluster $n$, respectively. According to the solutions of the knapsack problems \cite{WJ}, and the caching storage constraints in (\text{{\ref{P-0}c}}), (\text{{\ref{P-0}d}}), and \eqref{x_nfConstraint}, we have:}
\setcounter{equation}{21}
\begin{multline} \label{T^c}
T^\text c =\sum\nolimits_{n \in \mathcal N}\sum\nolimits_{m \in {{\cal M}_n^\text{c}}}  { { {\sum\nolimits_{f = 1}^{{K_n}} {\lambda _m} {p_{{nf}}^\text{o}} L} }}
+\\
\sum\nolimits_{m \in {{\cal M}}}{\sum\nolimits_{f = 1}^K {{\lambda _m}p_{mf}^\text{o}L}}.
\end{multline}
Define
\begin{equation}
{{T_m} =\sum\nolimits_{f = 1}^K {{\lambda _m}p_{mf}^\text{o}L}},
\end{equation}
\begin{equation}
{{T_n^\text{i}} = \sum\nolimits_{m \in {{\cal M}_n^\text{c}}} { {{\lambda _m}\left( {\sum\nolimits_{f = 1}^{{K_n}} {p_{{nf}}^\text{o}}  - \sum\nolimits_{f = 1}^K {p_{{mf}}^\text{o}} } \right)L}}},
\end{equation}
\begin{equation}
{T^\text{i}} =\sum\nolimits_{n\in \mathcal N} T_n^\text{i},
\end{equation}
where ${{T_m} }$ denotes the offloaded traffic at F-AP $m$ through fetching files that are cached in its own storage space,
${{T_n^\text{i}} }$ denotes the incremental offloaded traffic of cluster $n$,
and ${{T^\text{i}} }$ denotes the incremental offloaded traffic of all the $N$ clusters.
Then, \eqref{T^c} can be further expressed in an equivalent form as follows {\cite{XCui}}:
\begin{align}
T^\text c &={T^\text{i}} + \sum\nolimits_{m \in \mathcal M} {T_m}. \label{TmKnap}
\end{align}
It can be readily seen that the second item in the right hand side of \eqref{TmKnap} is unaffected by the clustering strategy, and that maximizing ${T^{\text c}}$  under the constraints in (\text{{\ref{P-0}a}})-(\text{{\ref{P-0}d}}) is equivalent to maximizing  $T^{\text i}$  under the constraints in (\text{{\ref{P-0}a}})-(\text{{\ref{P-0}b}}). Therefore, we can reformulate the clustering subproblem to maximize $T^{\text c}$ as follows:
\begin{align}\label{P-1}
&{\mathop {\text{max} }\limits_{{\left\{ {{\cal M}_n^{\text c}} \right\}_{n \in \mathcal N},{{\cal M}^{\text n}}}}}  { {T^{\text{i}} } } \\
{\text{s.t.}}\
&{(\text{{\ref{P-0}a}}), (\text{{\ref{P-0}b}}).} \tag{\ref{P-1}a}\nonumber
\end{align}

\setcounter{TempEqCnt}{\value{equation}}
\setcounter{equation}{31}
\begin{figure*}[!t]
\begin{multline} \label{T_d_trans}
{T^\text{d}}=\sum\nolimits_{n \in \mathcal N} {\sum\nolimits_{m \in {\cal M}_n^\text{c}} {\sum\nolimits_{f \in {\cal F}_n^{\text c} } {{\lambda _m}{p_{mf}}\left[ {1 - \prod\nolimits_{m' \in {\cal S}_m^2 \cup {\cal S}_m^3} {{\left( {1 - {x_{m'f}}} \right)} } } \right]L} } }
+\\
\sum\nolimits_{m \in {{\cal M}^\text{n}}} {\sum\nolimits_{f \in  {{\cal F}_{m}^{\text n}}} {{\lambda _m}{p_{mf}}\left[ {1 - \prod\nolimits_{m' \in {\cal S}_m^2 } {\left( {1 - {x_{m'f}}} \right)} } \right]L} }.
\end{multline}
\hrulefill
\end{figure*}
\setcounter{equation}{\value{TempEqCnt}}

Solving the above optimization problem, we can obtain the clustered and nonclustered F-AP sets.
Let ${\cal F}_n^{\text c} $ and ${\cal F}_m^{\text n}$ denote the set of $K_n$ most popular files in cluster $n$ and the set of $K$ most popular files at F-AP $m \in {\cal M}^{\text n}$, respectively.
Then, they can be expressed as follows:
\begin{multline}
{\cal F}_n^{\text c} = \left\{f \left| p_{n1}^{\rm{o}} \ge p_{n2}^{\rm{o}} \ge \cdots \ge p_{nf}^{\rm{o}} \ge  \cdots \ge p_{n{K_n}}^{\rm{o}} \right. \right\}, \\  n \in \mathcal N, \label{F_n}
\end{multline}
\begin{multline}
{\cal F}_m^{\text n} = \left\{f \left| p_{m1}^{\rm{o}} \ge p_{m2}^{\rm{o}} \ge \cdots \ge p_{mf}^{\rm{o}} \ge \cdots \ge p_{mK}^{\rm{o}}\right. \right\},\\
m \in {\cal M}^{\text n}. \label{F_m}
\end{multline}
Correspondingly, the caching decisions $\left\{x_{nf},x_{mf} \left| f \in {\cal F}, \right.\right.$ $ \left.  n \in \mathcal N, m \in {\cal M}^{\text n}  \right\}$ through maximizing $T^{\text c}$ can be expressed as follows:
\begin{equation} \label{x_result}
{x_{nf}} = \left\{ {\begin{array}{*{20}{c}}
{1,}&{f \in {\cal F}_n^{\text c},}\\
{0,}&{f \in {\cal F}  \backslash  {\cal F}_n^{\text c},}
\end{array}} \right. 
\end{equation}
\begin{equation} \label{x_result1}
{x_{mf}} = \left\{ {\begin{array}{*{20}{c}}
{1,}&{f \in {\cal F}_m^{\text n},}\\
{0,}&{f \in {\cal F}  \backslash {\cal F}_m^{\text n}.}
\end{array}} \right.
\end{equation}
Substitute \eqref{x_result} and \eqref{x_result1} into (\ref{T_s2s3}). Then, $T^{\text d}$ can be expressed  in an equivalent form in  \eqref{T_d_trans}   as shown at the top of next page. Therefore, we can reformulate the content placement subproblem to minimize $T^{\text d}$ as follows:
\stepcounter{equation}
\begin{align}\label{P-2}
&{\mathop {\text{min} }\limits_{x_{mf}}}  {{T^\text{d}}} \\
{\text{s.t.}}\
&{(\text{{\ref{P-0}a}}), (\text{{\ref{P-0}b}}),(\text{{\ref{P-0}c}}),(\text{{\ref{P-0}d}}).} \tag{\ref{P-2}a}\nonumber
\end{align}
For convenience, a summary of major notations is presented in Table I.

\section{Proposed Graph-based Cooperative Caching Scheme}

In the previous Section, we have transformed the challenging cooperative caching optimization problem into a clustering  subproblem and a content placement subproblem.
The clustering subproblem in (\ref{P-1}) and the content placement subproblem in (\ref{P-2}) fall into the scope of combinatorial programming {\cite{Ashraf, XCui}}. A brute force approach is generally required to obtain the globally optimal solution of each subproblem. However, such an approach has an exponential complexity w.r.t. the number of F-APs and the number of disjoint cluster sets
or the sizes of popular file sets ${\cal F}_n^{\text c} $ and ${\cal F}_m^{\text n}$.
Although its computational complexity is indeed reduced compared to the original dynamic programming approach,
it is still computationally impracticable even for a small size network.
By mapping each F-AP as one vertex in a graph, the candidate cluster can be represented by its subgraph \cite{Zhou}.
By mapping each obtained subgraph as the vertex in a new graph, the disjoint cluster sets can be represented by an independent subset of the vertex set of this new graph. According to graph theory \cite{Bondy}, all the subgraphs of a graph and the independent subset of the vertex set of a graph can be obtained in polynomial time complexity. Correspondingly, the clustering subproblem can be solved in polynomial time complexity.
Furthermore, by mapping each pair of cooperative F-APs which are not in the same cluster as two vertices that are connected by one edge in a graph,  all the edges  can be traversed to control the cached files at the corresponding paired F-APs and the duplicate cached files can then be eliminated.
According to graph theory \cite{Bondy}, all the edges in a graph can be found in polynomial time complexity. Correspondingly, the content placement subproblem can also be solved in polynomial time complexity.
Therefore, we commit to an effective graph-based approach to solve the clustering subproblem and the content placement subproblem, respectively.

\subsection{Proposed Graph-based Clustering Approach}

\subsubsection{Description of the Proposed Approach}
\begin{figure}[!t]
\centering 
\includegraphics[width=0.45\textwidth]{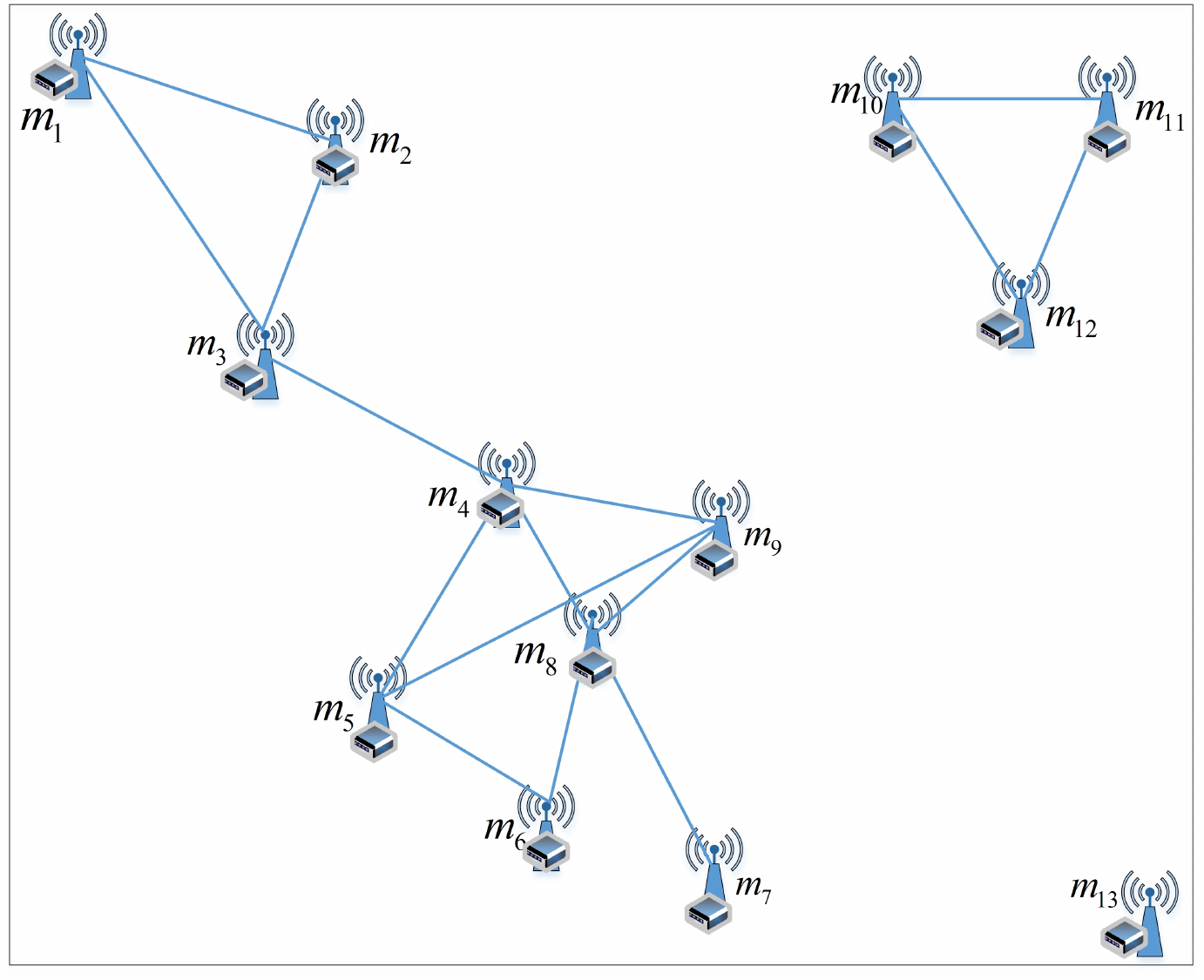}
\caption{Illustration of a node graph including thirteen vertices.}
\label{nodegra}
\end{figure}

In our proposed graph-based clustering approach,
firstly, all the considered $M$ F-APs are checked to determine which pair satisfies the constraints in (\ref{P-1}a). 
It is already known that F-APs with appropriate distance and load difference from each other are more likely to cooperate together \cite{Ashraf}.
Then, according to the checking results, the node graph denoted by ${{\cal G}^{\text n}} = ({\cal M},{\cal E})$ is constructed, whose vertex set denoted by $\cal M$  is the F-AP set and whose edge set denoted by $\cal E$ reflects the distance and load difference among the F-APs.
In ${{\cal G}^{\text n}} $, two vertices are connected through an edge if their representing F-APs can cooperate with each other.
Note that one  subgraph of ${{\cal G}^{\text n}}$,  any vertex of which can connect through an edge with a certain vertex in the same subgraph, represents one cluster which consists of a certain number of  cooperative F-APs,
and one complete subgraph of ${{\cal G}^{\text n}}$, any two vertices of which can connect through an edge,  essentially represents
one candidate cluster of the optimization problem in \eqref{P-1} whose cluster members can cooperative with each other.
We point out here that there may exist a certain vertex not belonging to any subgraph of ${{\cal G}^{\text n}}$, which means that  its representing F-AP is nonclustered.
For illustration, a node graph with thirteen vertices  as shown in  Fig. \ref{nodegra} is taken for example.
According to the above descriptions, seeking candidate clusters is equivalent to searching for  complete subgraphs in ${{\cal G}^{\text n}}$.
The algorithm of searching for  complete subgraphs will be presented in detail in Section IV-A-2.

Let ${\cal H} = \{ {h_1},{h_2}, \cdots ,{h_n}, \cdots ,{h_{N'}}\}$ denote the complete subgraph set
that has been obtained through the above searching algorithm, where $N'$ denotes the number of the complete subgraphs so obtained.
It is clear that $\left\{ {{\cal M}_n^{\text c}} \right\}_{n = 1}^N \subseteq {\cal H}$.
Then, a weighted graph denoted by ${{\cal G}^{\text w}} = ({\cal H},{\cal B}, \boldsymbol w)$ can be constructed,
where $\cal H$ denotes the vertex set,
$\cal B$ denotes the edge set, and $\boldsymbol w$ denotes the weight vector corresponding to the vertices of ${{\cal G}^{\text w}}$ whose elements are set to be the incremental offloaded traffic of their corresponding complete subgraphs, i.e., {${\left[ \boldsymbol w \right]_n} = \mathop T\nolimits_n^{\text i} $}.
In ${{\cal G}^{\text w}}$, two vertices are connected through an edge if their representing complete subgraphs have a certain identical vertex.
It is known from graph theory that an independent or stable set is a set of vertices in a graph, {no two of which are adjacent \cite{Bondy}.
Then, the independent subset of ${\cal H}$ certainly satisfies the constraint in \eqref{eq5}.
Correspondingly, the objective cluster sets of the optimization problem in \eqref{P-1} can be readily obtained by searching for the equivalent max-weight independent subset of ${\cal H}$ of  the corresponding weighted graph ${{\cal G}^{\text w}}$.
The max-weight independent subset of ${\cal H}$ can be obtained by solving  {a 0-1 integer programming problem,
which will be  presented in detail in Section IV-A-3.

Remark here that we map one cluster to one complete subgraph,
which guarantees proper-sized clusters and avoids unnecessary intra-cluster signaling overhead,
instead of one connected subgraph as in \cite{Zhou},
{which tends not to constrain the cluster size.}

\subsubsection{Searching for Complete Subgraphs}
We propose to search for maximal complete subgraphs to find all the possible complete subgraphs.
It is known from \cite{Bondy} that any complete subgraph must belong to a maximal complete subgraph
and it is more difficult  to find complete subgraphs through direct searching than through indirect searching for maximal complete subgraphs.
We propose to exploit the adjacency table of each vertex in the node graph ${{\cal G}^{\text n}}$ to search for maximal complete subgraphs.
For $m \in {\cal M}$, let ${{\cal T}_m} = \left\{ m \right\} \cup \left\{ {{m'}\left| {m' \in {\cal M},m' > m,\left( {m',m} \right) \in {\cal E}} \right.} \right\}$ denote the adjacency table of vertex $m$ of ${{\cal G}^{\text n}}$,
and ${L_m}$ denotes the table size of ${{\cal T}_m}$.
If {${L_m} = 1$} or ${{\cal T}_m} \subseteq {{\cal T}_{m'}} $ for $ m, m' \in {\cal M}$ and $m' <m$,
it is unnecessary to search for a maximal complete subgraph in ${{\cal T}_m}$.
Remove all the unnecessary or redundant  adjacency tables
and sort the remaining in descending order denoted by  ${\cal T} _{ m }^ {\text o}$ according to their table sizes.
Let ${{\cal T}}$ denote the set of the reordered adjacency tables of ${{\cal G}^{\text n}}$.
Remove any vertex that does not connect with all the other vertices in ${\cal T} _{ m }^ {\text o}$.
Then, the remaining vertices in ${{\cal T}_m ^ {\text o}}$ form a maximal complete subgraph.
Let ${{\cal G}^{\text m}}$ denote the set of maximal complete subgraphs.
The detailed description of our proposed algorithm of  searching for maximal complete subgraphs is presented in Algorithm \ref{alg-1}.
After maximal complete subgraphs are found, all the possible complete subgraphs can be readily obtained.

\begin{algorithm}[!t]
	\renewcommand{\algorithmicrequire}{\textbf{Input:}}
	\renewcommand{\algorithmicensure}{\textbf{Output:}}
	\caption{Searching for Maximal Complete Subgraphs}
	\label{alg-1}
	\begin{algorithmic}[1]
		\Require ${{\cal G}^ \text n}$
        \Ensure ${{\cal G}^{\text m}}$            		
        \For {each ${\cal T}_m^{\text o} \in {{\cal T}}$}
        \State Initialize $i=0$,  ${\cal T}_{i}^{\text t}=\varnothing$, ${\cal T'} = {\cal T}_m^{\text o}$;
        \For {each ${j} \in {\cal T'}$}
        \If {${\cal T}_m^{\text o}$ or ${\cal T}_{i'}^ \text t \in \left\{ {{\cal T}_{i'}^\text t} \right\}_{i' = 0}^{i - 1}$ contains both vertex \\
                \qquad \quad $j$ and nonadjacent vertices of $j$}
        \State ${\cal T}_i^\text t = {\cal T}_m^\text o - {{\cal T}_j}$ or ${\cal T}_i^\text t = {\cal T}_{i'}^\text t - {{\cal T}_j}$;
        \State Remove all the nonadjacent vertices of $j$ \\
        \qquad \quad \quad \ \ from the corresponding ${\cal T}_m^{\text o}$ or ${\cal T}_{i'}^\text t$, set \\
        \qquad \quad \quad \ \ $i = i + 1$;
        \EndIf
        \EndFor
        \State ${{\cal G}^{\text m}} = {{\cal G}^{\text m}} \cup {{\cal T}_m^{\text o}} \cup \left\{ {{\cal T}_{i'}^\text t} \right\}_{i' = 0}^{i-1}$
        \EndFor
	\end{algorithmic}
\end{algorithm}

\subsubsection{Searching for Max-Weight Independent Subset}

According to the construction of the weighted graph ${{\cal G}^{\text w}}$,
two vertices in $\cal H$ are adjacent and there exists an edge between them if their representing  candidate cluster sets have some identical elements.
Let ${\boldsymbol x}$ denote the binary indicating vector for the the vertices in $ \cal H$ with ${[\boldsymbol x]_n} = 1$
if the candidate cluster set represented by the vertex $h_n$ belongs to the objective disjoint cluster sets of the original optimization problem in \eqref{P-1}
and ${[\boldsymbol x]_n} = 0$ otherwise.
If the vertices $h_n$ and $h_{n'}$ can be connected through an edge $\left( {{h_n},{h_{n'}}} \right) \in {\cal B}$,
the relationship ${[\boldsymbol x]_n}  [\boldsymbol x]_{n'} = 0$ should be satisfied.

According to the above description, the original optimization problem in  \eqref{P-1} can be transformed into the following  0-1 integer programming problem,
\begin{align}\label{eq10}
&\mathop {\max }\limits_{\boldsymbol x} { {\boldsymbol w ^T {\boldsymbol x}} }\\
{\text{s.t.}}\ & {{[\boldsymbol x]_n} \in \{ 0,1\}}, \forall h_n \in \cal H, \tag{\ref{eq10}a}\\
&{{{[\boldsymbol x]_n}  {[\boldsymbol x]_{n'}} = 0,\forall \left( {{h_n},{h_{n'}}} \right) \in {\cal B}}}. \tag{\ref{eq10}b}
\end{align}
The above optimization problem can be solved by linear programming only if its linear relaxation is tight and has a unique integral solution. However, the above two conditions  are hard to be satisfied \cite{Sanghavi}.
Actually, the optimization problem in \eqref{eq10} is a classical problem that maximizes a submodular set function and can often be solved by greedy algorithms \cite{JV}.
Considering that traditional greedy algorithms cannot take full advantage of the specific constraints in (\ref{eq10}a)-(\ref{eq10}b), we then propose a more effective greedy algorithm.
Let  ${{\cal G}_n}$ denote the independent subset of $\cal H$ and
 $w_n$ denote the sum weight of all the vertices in ${{\cal G}_n}$.
Each time   move one vertex with the largest weight from $\cal H$ to  ${{\cal G}_n}$
and remove its adjacent vertices from $\cal H$.
Repeat the above step until $\cal H$ is empty.
The independent subset ${{\cal G}_n}$ so obtained with the maximum sum weight $w^{\text n}$ is just the max-weight independent subset denoted by ${\cal G}^{\text o}$ that we are searching for.
The detailed description of our proposed greedy algorithm of searching for the max-weight independent subset is presented in Algorithm \ref{alg-2}.

\begin{algorithm}[!t]
	\renewcommand{\algorithmicrequire}{\textbf{Input:}}
	\renewcommand{\algorithmicensure}{\textbf{Output:}}
	\caption{Searching for Max-Weight Independent Subset}
	\label{alg-2}
	\begin{algorithmic}[1]
		\Require ${\cal G} ^ \text w$
        \Ensure {${{\cal G}^{\text o}}$}
        \State Initialize {$w^{\text o}=0$, ${{\cal G}^{\text o}} = \varnothing $};	
        \For {each ${h_n} \in {\cal H}$}
        	\State {Initialize ${{\cal H'}} = {\cal H}$,} ${{\cal G}_{n}} = \left\{ {{h_n}} \right\}$, $w_{n} = [\boldsymbol w ]_n$;
        	\State Remove all the adjacent vertices of $h_n$ from {${\cal H'}$};
        	\While {${\cal H'} \ne \varnothing $}
                \State Find the vertex with the largest weight from \\
                \quad \quad \quad  {${\cal H'}$} denoted by $h_{{n^{\text max}}}$;
       			\State ${{\cal G}_n} = {{\cal G}_{n}} \cup \left\{ {{h_{{n^{\text max}}}}} \right\}$, $w_{n} = w_{n} + [\boldsymbol w ]_{n^{\text max}}$;
        		\State Remove all the adjacent vertices of $h_{n^{\text max}}$ \\
                  \quad \quad \quad  from {${\cal H'}$};
        	\EndWhile
        	\If {${w^{\text o}} < w^{\text n}$}
        		\State ${w^{\text o}} = w_{n}$, ${{\cal G}^{\text o}} = {{\cal G}_{n}}$.
        	\EndIf
        \EndFor
	\end{algorithmic}
\end{algorithm}

In traditional greedy algorithms \cite{JHoepman, SBasagni}, the vertex with the largest weight is generally chosen as the initial vertex to search  for the max-weight independent subset. In contrast, we set $N'$ outer loops in Algorithm \ref{alg-2}. Correspondingly, each vertex in $\cal H$ has chance to be the initial vertex to constitute an independent subset. Therefore, the $N'$ outer loops in Algorithm \ref{alg-2} guarantee to find the max-weight independent subset of the vertex set of the weighted graph.

To further illustrate the above issue, take a weighted graph with nine vertices as shown in Fig. \ref{local} for example. In the weighted graph, the vertices are divided into three groups according to their weights,
the weight of each vertex in the first group is larger than that in the second and third groups, and the weight of each vertex in the second group is larger than that in the third group. Assume vertex 1 has the largest weight among the nine vertices.
In traditional greedy algorithms, vertex 1 will be chosen as the initial vertex.
Then, the output max-weight independent subset will be $\{ 1,2,3\}$.
However, in Algorithm \ref{alg-2}, vertex 4 is also allowed to be the initial vertex. Then, the output max-weight independent subset will be $\left\{ {4,5,6,7,8} \right\}$ if its sum weight is larger than that of $\{ 1,2,3\}$.
It can be readily seen that the sum weight of all the vertices in the obtained independent subset will not be the maximum if the initial vertex is not selected properly.
Therefore, the $N'$ outer loops in Algorithm \ref{alg-2} can indeed guarantee to find the max-weight independent subset.

After the max-weight independent subset ${\cal G}^{\text o}$ is found, the clustered sets and nonclustered set can be determined. Then, the set of popular files ${\cal F}_n^{\text c}$ for {$n \in {\cal N}$} in  cluster $n$  and the set of popular files ${\cal F}_m^{\text n}$ for {$m \in {\cal M}^{\text n}$} at the nonclustered F-AP  $m$ can be determined according to \eqref{F_n} and \eqref{F_m}, respectively.

\begin{figure}[!t]
\centering 
\includegraphics[width=0.45\textwidth]{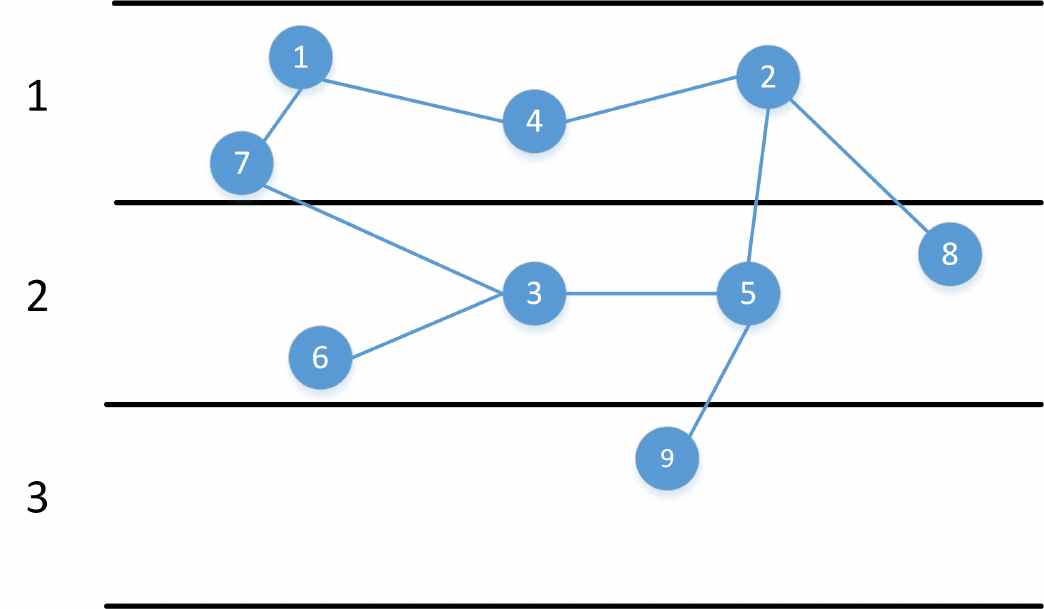}
\caption{Illustration of a weighted graph with nine vertices.}
\label{local}
\end{figure}

\subsection{Proposed Graph-based Content Placement Approach}

\subsubsection{Description of the Proposed Approach}
In our proposed graph-based content placement approach,
firstly, we find the complete subgraphs corresponding to the elements in the obtained max-weight independent subset ${\cal G}^{\text o}$, and remove the edges in these complete subgraphs. Utilizing the vertices and  the remaining edges  in the node graph ${\cal G}^{\text n}$,
we propose to construct a redundancy graph denoted by ${\cal G}^{\text r}=\left({\cal M},{\cal E}^{\text r}\right)$, where ${\cal M}$ denotes its vertex set and ${\cal E}^{\text r}$ denotes its edge set reflecting the cache redundancy among cooperative F-APs.
Let $e = \left\{ {m,m'} \right\} \in {\cal E}^{\text r}$ denote the edge that connects vertex $m$ and vertex $m'$, and
${\cal F}_{e}^\text d$ denote the set of duplicate popular files in the obtained popular file set of the cooperative F-APs  corresponding to the vertices connected by edge $e$. If edge $e$ connects $m \in {{{\cal M}_n^{\text c}}}$ and $ m' \in {\cal M}^\text n$, then we have: ${\cal F}_e^\text d = {\cal F}_n^{\text c} \cap {\cal F}_m^{\text n}$. If edge $e$ connects $m \in {{\cal M}_n^{\text c}}$ and $ m' \in {{\cal M}_{n'}^{\text c}}$, then we have: ${\cal F}_e^{\text d} = {\cal F}_n^{\text c} \cap {\cal F}_{n'}^{\text c}$, whose size  may exceed $K$, i.e., the storage size of each F-AP. Correspondingly, only a portion of the duplicate popular files in ${\cal F}_{e}^\text d$ will indeed cause cache redundancy.
Furthermore, when edge $e$ connects $m \in {{\cal M}_n^{\text c}}$ and edge $e'$  connects $ m'\in {{\cal M}_{n}^{\text c}}$ with $e' \ne e$ and $m' \ne m$, ${\cal F}_{e}^\text d$ and ${\cal F}_{e'}^\text d$ may contain duplicate files. Correspondingly, only a portion of the duplicate popular files in ${\cal F}_{e}^\text d$ and ${\cal F}_{e'}^\text d$ will indeed cause cache redundancy.
Therefore, we propose to separate ${\cal F}_{e}^\text d$ to determine the duplicate files that will indeed cause cache redundancy at edge $e$.
The process of separating ${\cal F}_{e}^\text d$  will be  presented in detail in Section IV-B-2.

Then, we propose to enhance the caching decisions to control the caching locations of the duplicate popular  files and ensure that each duplicate popular file is  cached only once between each pair of cooperative F-APs.
After determining the caching locations for all the duplicate popular files, the remaining storage space of each F-AP is filled by the rest files according to their request probability. The process of caching-decision enhancement will be  presented in detail in Section IV-B-3.

\subsubsection{Separate the Set of Duplicate Popular Files}

Let ${\cal T}_m$ denote the adjacency table of vertex $m$ of ${\cal G}^{\text r}$. Sort all the adjacency tables in descending order according to their table sizes. Let  ${\cal T}$ denote the set of the reordered adjacency tables  of ${\cal G}^{\text r}$, and ${\cal T}_m^{\text o}$ denote the $m$th adjacency table in ${\cal T}$.
Let $K_m$ denote the size of the remaining storage space of the F-AP corresponding to vertex $m$. Initialize $K_m=K $.
Let ${\cal F}_m^\text i$ denote the intersection of the sets of duplicate popular files at all the edges that connect vertex $m$ and its adjacency vertices in ${\cal T}_m^{\text o}$. Then, it can be  expressed as follows:
\begin{equation}
{\cal F}_m^\text i = \bigcap\nolimits_{ e = \{m,  m' \}, m' \ne m, m'  \in {\cal T}_m^{\text o}} {{\cal F}_{e}^\text d}, \ m \in \mathcal M.
\end{equation}
Let ${\cal F}_{e}^{\text r}$ denote the set of  
files that will indeed cause cache redundancy}
at edge $e$ after separating ${\cal F}_{e}^{\text d}$.
If $\left| {\cal F}_m^\text i \right| \ge K_m$,  ${\cal F}_{e}^{\text r}$ will be constituted by the random $K_m$ files in ${\cal F}_m^\text i$. Otherwise, ${\cal F}_{e}^{\text r}$ will be constituted by the random ${{(K_m- \left| {\cal F}_m^\text i \right| )} \mathord{\left/ {\vphantom {{{K_m}} ({\left| {{\cal T}_m^{\text o}} \right|}-1)}} \right. \kern-\nulldelimiterspace} ({\left| {{\cal T}_m^{\text o}} \right|}-1)}$  files in ${\cal F}_{e}^\text d \backslash {\cal F}_m^\text i$ and all the files in ${\cal F}_m^\text i$. Once ${\cal F}_{e}^{\text r}$ is determined, update $K_{m'}=K_{m'}- \left| {\cal F}_{e}^{\text r} \right|$ for vertex $m'  \in {\cal T}_m^{\text o}$ and   update ${\cal F}_{e'}^\text d ={\cal F}_{e'}^\text d \backslash {\cal F}_{e}^{\text r}$ if edge $e' \in {\cal E}^{\text r}$ connects vertex $m'$.

\subsubsection{Enhance the Caching Decisions}

Let $\Delta {x_{mf}} \in \left\{ -1,0,1 \right\}$ denote the indicator of the caching-decision enhancement for file $ f \in \cal F$ at vertex $m \in {\cal M}$, where $\Delta {x_{mf}} = 1$ indicates that the F-AP corresponding to vertex $m$ is chosen as the caching location for file $f$, $\Delta {x_{mf}} = -1$ indicates that the F-AP corresponding to vertex $m$ is not allowed to cache file $f$ so as to eliminate redundancy, and $\Delta {x_{mf}} = 0$ indicates that the caching location for file $f$ has not been determined yet. Initialize ${\Delta {x_{mf}}}=0$ and set $K_m=K$.

Firstly, calculate the indicators of the caching-decision enhancements for file $f \in {\cal F}_e^{\text r}$. For each ${\cal T}_m^{\text o} \in {\cal T}$ and each $m'  \in {\cal T}_m^{\text o}$ with $m' \ne m$, find the files whose caching locations are at the F-AP corresponding to vertex $m$, and  forbid these files to be cached at the F-AP corresponding to  vertex $m'$. Then, the indicators of the corresponding caching-decision enhancements are set as follows:
\begin{equation}  \label{First-correction-S}
\Delta {x_{m'f}} = -1, \  f \in \left\{ f \left| \Delta {x_{mf}} = 1 \right. \right\} \cap {\cal F}_{e}^{\text r}.
\end{equation}
Update ${\cal F}_{e}^{\text r}$ by removing these files. Furthermore, find the files whose caching locations  {are not allowed to be} at the F-AP corresponding to vertex $m$, and choose the F-AP corresponding to vertex $m'$ as the caching locations for these files. Then, the indicators of the corresponding caching-decision enhancements are set as follows:
\begin{equation}
\Delta {x_{m'f}} = 1, \  f \in  \left\{ f \left| \Delta {x_{mf}} = -1 \right. \right\} \cap {\cal F}_{e}^{\text r}.
\end{equation}
Update $K_{m'}$ and ${\cal F}_{e}^{\text r}$ by removing these files. {If F-AP $m' \in {\cal M}_n^{\text c}$, update ${\cal F}_{n}^{\text c}$
by removing these files.} Let $T_{em}^\text{p}$ denote the possible offloaded traffic due to caching the remaining  files in ${{\cal F}_{e}^{\text r}}$ at the F-AP corresponding to vertex $m$. Then, it can be  expressed as follows:
\begin{equation} \label{T^p}
T_{em}^\text{p} = \sum\nolimits_{m' \in m \cup {{\cal S}_m}} {\sum\nolimits_{f \in {{\cal F}_{e}^{\text r}}} {{\lambda _{m'}}{p_{m'f}}L} }.
\end{equation}
Suppose $T_{em}^\text{p} \ge  T_{em'}^\text{p}$. Then, set the indicators of the corresponding caching-decision enhancements as follows:
\begin{equation} \label{First-correction-E}
\Delta {x_{mf}} = 1,\Delta {x_{m'f}} = -1, \  f \in {\cal F}_{e}^{\text r}.
\end{equation}
Update $K_{m}=K_{m}- \left| {\cal F}_{e}^{\text r} \right|.$
If F-AP $m \in {\cal M}_n^{\text c}$, update ${\cal F}_{n}^{\text c}$ by removing these files.

Secondly, calculate the indicators of the corresponding caching-decision enhancements for the remaining files in ${\cal F}_n^{\text c}$.  For each $m \in {\cal M}_n^{\text c}$,
find the files whose caching locations can be at the F-AP corresponding to vertex $m$. Then, set the indicators of the corresponding caching-decision enhancements as follows:
\begin{equation}\label{Second-correction-S}
\Delta {x_{mf}} = 1,  \   f \in \left\{ f \left| \Delta {x_{mf}} =0 \right. \right\} \cap {\cal F}_n^{\text c}.
\end{equation}
Update $K_m$ and  ${\cal F}_{n}^{\text c}$ by removing these files. For each $m \in {\cal M}_n^{\text c}$ which satisfies $K_m>0$,   randomly select $K_m$ files  from ${\cal F}_n^{\text c}$, and set the indicators of the corresponding caching-decision enhancements as follows:
\begin{equation}\label{Second-correction-E}
\Delta {x_{mf}} = 1,  \  f \in {\cal F}_n^{\text c}.
\end{equation}
Update $K_m$ and  ${\cal F}_{n}^{\text c}$ by removing these files.

Thirdly, calculate the indicators of the corresponding caching-decision enhancements at the vertices corresponding to nonclustered F-APs. For each $m \in {{\cal M}^\text n}$ which satisfies  $K_m>0$,   select $K_m$ most popular files  from   the uncached files at the F-AP corresponding to vertex $m$ and its cooperators according to their request probability, and set $\Delta {x_{mf}}$ as follows:
\begin{equation}\label{Final-correction}
\Delta {x_{mf}} = 1, \  f \in {\cal F} \backslash \left\{ {f\left| {\Delta {x_{m'f}} = 1, m' \in m \cup {{\cal S}_m}} \right.} \right\}.
\end{equation}

Finally, enhance the caching-decision for each $m \in {{\cal M} }$ as follows:
\begin{equation} \label{objective}
x_{mf} = \left\{ {\begin{array}{*{20}{c}}
{1,}&{\Delta {x_{mf} } = 1,}\\
{0,}&{\Delta {x_{mf} } \ne 1.}
\end{array}} \right.
\end{equation}

The detailed description of our proposed graph-based content placement algorithm is presented in Algorithm \ref{alg-3}.

\begin{algorithm}[!t]
	\renewcommand{\algorithmicrequire}{\textbf{Input:}}
	\renewcommand{\algorithmicensure}{\textbf{Output:}}
	\caption{Graph-based Content Placement Algorithm}
	\label{alg-3}
	\begin{algorithmic}[1]
		\Require ${\cal F}_m^{\text n}$, ${\cal F}_n^{\text c}$, ${\cal G}^{\text r}$
        \Ensure $x_{mf}$
        \State  Calculate ${\cal F}_e^{\text r}$ for each edge $e$;       		
        \For {each  ${\cal T}_m^{\text o} \in {{\cal T}}$}
        \For {each $m' \in {\cal T}_m^{\text o}$ with $m' \ne m$}
        \State Determine $\Delta {x_{mf} }$ and $\Delta {x_{m'f} }$ according to (\ref{First-correction-S})-
        \State (\ref{First-correction-E}), and update the corresponding ${\cal F}_e^{\text r}$, ${\cal F}_n^{\text c}$, $K_m$,
        \State and $K_m'$;
        \EndFor
        \EndFor
        \For {each $n \in \mathcal N$}
        \For {each $m \in {\cal M}_n^{\text c}$}
        \State Determine $\Delta {x_{mf} }$ according to (\ref{Second-correction-S}), and update
        \State the corresponding $K_m$ and ${\cal F}_n^{\text c}$;
        \EndFor
        \For {each $m \in {\cal M}_n^{\text c}$}
        \State Determine $\Delta {x_{mf} }$ according to (\ref{Second-correction-E}), and update
        \State the corresponding $K_m$ and ${\cal F}_n^{\text c}$;
        \EndFor
        \EndFor
        \For {each $m \in {{\cal M}^\text{n}}$}
        \State Determine $\Delta {x_{mf} }$ according to (\ref{Final-correction});
        \EndFor
        \For {each $m \in {{\cal M} }$}
        \State Set $x_{mf}$ according to (\ref{objective}).
        \EndFor
	\end{algorithmic}
\end{algorithm}

\subsection{Complexity Analysis}
Let ${\bar L}$ denote the average size of the adjacency tables of all the vertices in the node graph $\cal G^ \text n$.
Then, the computational complexity of searching for maximal complete subgraphs in Algorithm \ref{alg-1} is  ${\mathcal O}(M\bar L)$.
Furthermore, the computational complexity of obtaining all the complete subgraphs is ${\mathcal O}(P{{\bar V}})$, where  $P$ denotes the number of the maximal complete subgraphs that have been found,
and  {${{\bar V}}$} denotes the average vertex number of all the complete subgraphs.
Besides, the computational complexity of searching for the max-weight independent subset in Algorithm \ref{alg-2} is ${\mathcal O}(N'{N})$.
Therefore, the computational complexity of the proposed graph-based clustering approach is  ${\mathcal O}(M\bar L + P{{\bar V}} + N'{N})$.

Let $\delta $ denote the maximum degree of the redundancy graph. The computation complexity of the proposed graph-based content placement algorithm is ${\mathcal O}(M\delta  + 2M)$.

In summary,  the computational complexity of the proposed graph-based cooperative caching scheme is  {${\mathcal O}(M\bar L + P{{\bar V}} + N'{N} + M\delta + 2M)$}.
By considering  ${\bar L< M}$, ${{\bar V}} < M$, $N<M$, and $\delta <M$,
the computation complexity of the proposed graph-based cooperative caching scheme is ${\mathcal O}(M^2 + PM + N'M)$ for the worst case.
It is obviously lower than that of  ${\mathcal O}(M^3KF^2)$ in \cite{KShanmugam} and ${\mathcal O}(M^4K+MKF)$ in \cite{RWang} by taking $M \ll F, P< F,$ and  $N'< F$ into account.

\section{Simulation Results}

In this section, the performance of the proposed graph-based cooperative caching scheme is evaluated via simulations.
In the simulations,  the request probability at each F-AP is generated from the global request probability which follows Zipf distribution with the skewness parameter $z$.\footnote{Let $p_f$ denote the global request probability of file $f$. Assume that the global request probability and the request probability at the considered $M$ F-APs have the following relationship:
${p_f} = \sum\nolimits_{m \in {\cal M}} {{w_m}{p_{mf}}} $ \cite{Li}.}
Unless otherwise stated, the system parameters are set as follows:
$z=0.6$, $M=10$, $F=5000$, $K=250$, $L=2$ Gb. 
We choose the locally popular caching (LPC) scheme and the globally popular caching (GPC) scheme as two  baselines \cite{JL}. In the LPC scheme,  the most $K$ popular files are cached at each F-AP based on the local request probability, and neighboring F-APs can cooperate with each other. In the GPC scheme, the most $K$ popular files are cached at each F-AP based on the global request probability, and neighboring F-APs cannot cooperative with each other.

\begin{figure}[!t]
\centering 
\includegraphics[width=0.48\textwidth]{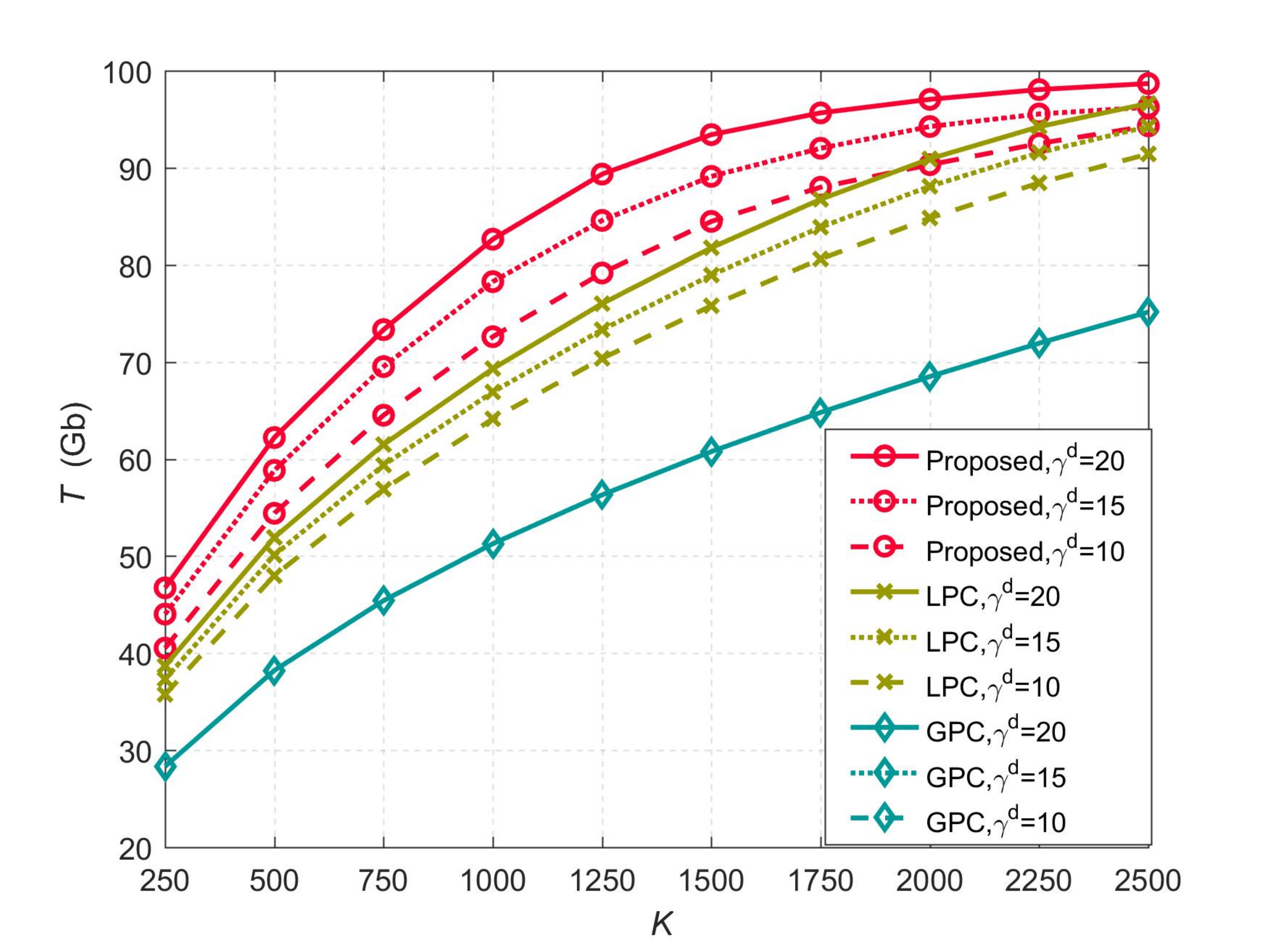}
\caption{The offloaded traffic $T$ versus the storage size $K$ of each F-AP under different distance thresholds ${\gamma ^{\text d}}$.}
\label{result4}
\end{figure}

In Fig. \ref{result4}, we show the  offloaded traffic $T$ of our proposed  scheme  and the two baselines versus the storage size $K$ of each F-AP with different distance threshold ${\gamma ^{\text d}}$.
It can be observed that the  offloaded traffic of all the three schemes increases with the storage size.
It can also be observed that the performance of the proposed scheme is superior to  that of the baselines.\footnote{Clearly, a centralized approach has been assumed in this paper.
This would certainly incur the necessary signaling overhead, and the impact of such overhead will be an interesting issue for future research.}
The reason is that the proposed scheme improves clustering and reduces the repetitive and redundant storage  of files. Correspondingly, more user requests can be satisfied locally  compared with the baselines.
Furthermore, the   offloaded traffic of the proposed and LPC schemes increases with distance threshold ${\gamma ^{\text d}}$, and ${\gamma ^{\text d}}$ has a greater influence on the performance of the proposed scheme.
The reason is that as ${{\gamma ^\text{d}}}$ becomes larger, the constraints of the clustering subproblem in our proposed scheme will be relaxed,  the cluster size will become larger, more F-APs can  cooperate with each other, and more files can then be successfully cached locally.

\begin{figure}[!t]
\centering 
\includegraphics[width=0.48\textwidth]{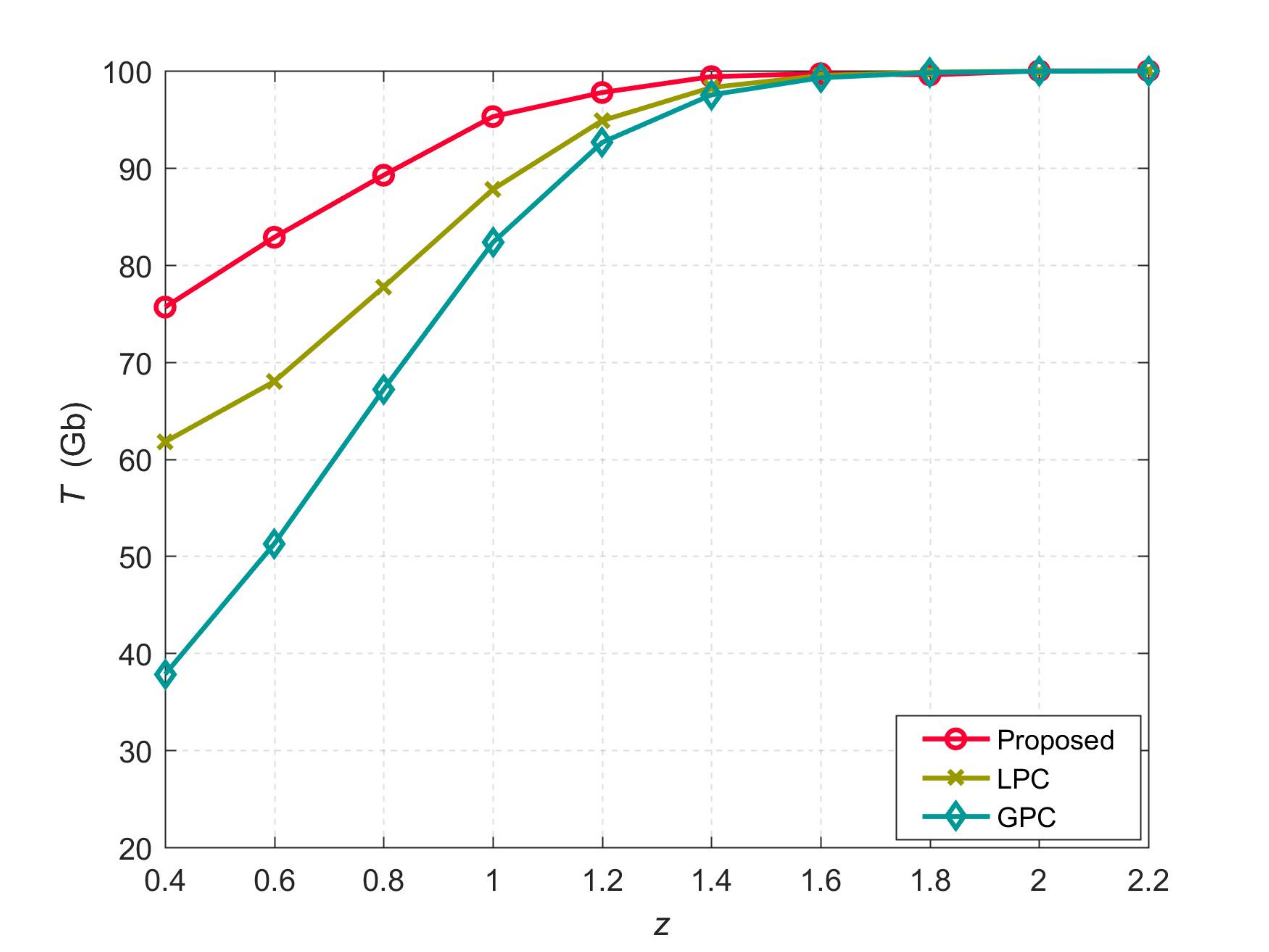}
\caption{The offloaded traffic $T$ versus  the skewness parameter $z$ of Zipf distribution with ${\gamma ^{\text d}}=20$ and $ K=1000$.}
\label{result5}
\end{figure}

In Fig. \ref{result5}, we show the offloaded traffic $T$ of our proposed  scheme  and the two baselines versus the skewness parameter $z$ of Zipf distribution with ${\gamma ^{\text d}}=20$ and $ K=1000$.
It can be observed that the   offloaded traffic of all the three schemes increases with $z$.
The reason is that as $z$ becomes larger, the most popular files will concentrate in a fewer files and more traffic can then be offloaded.
It can also be observed that the performance of the proposed scheme is superior to that of the baselines for all $z$.

\begin{figure}[!t]
\centering 
\includegraphics[width=0.48\textwidth]{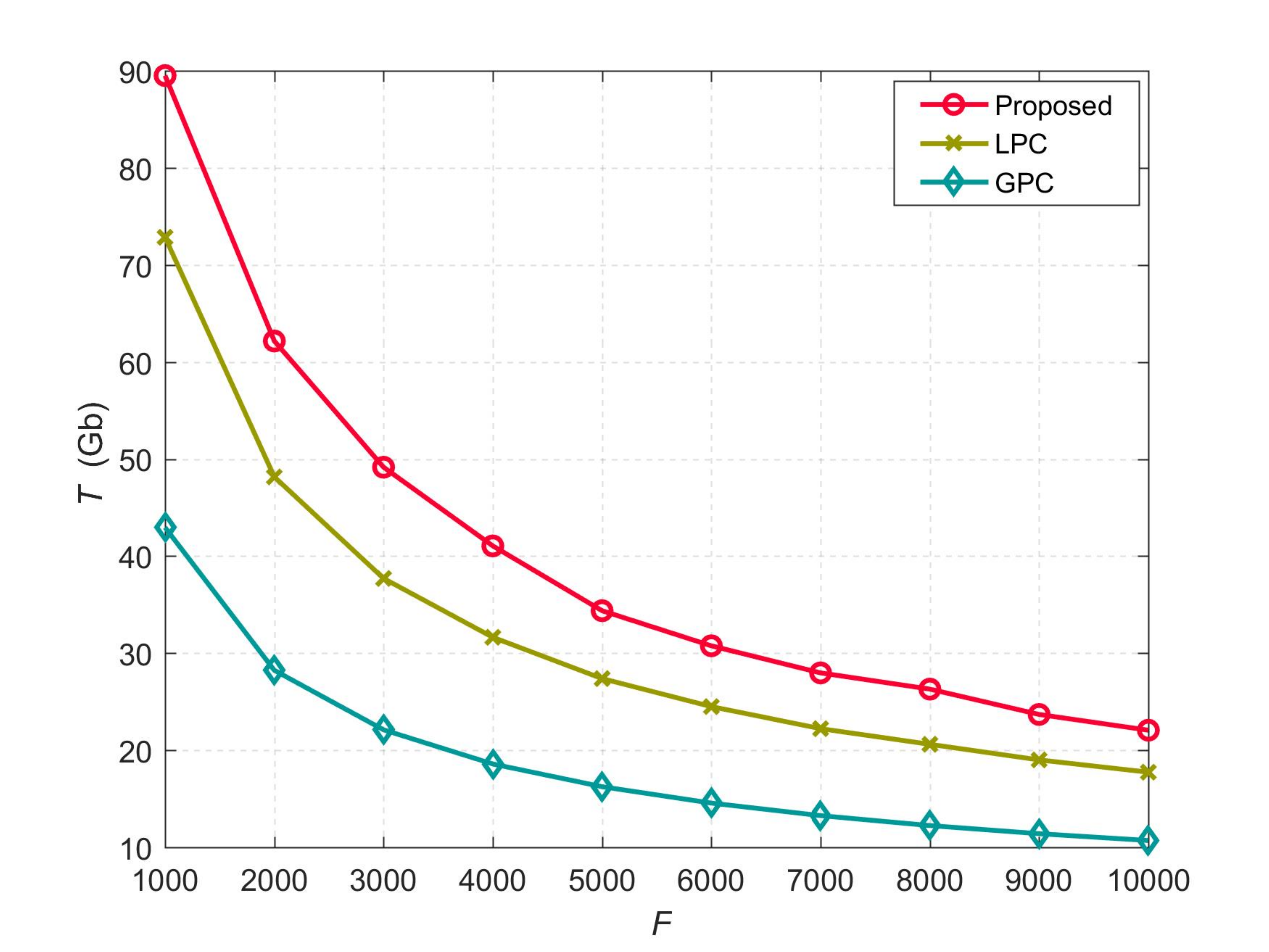}
\caption{The offloaded traffic $T$ versus the content library size $F$ with ${\gamma ^{\text d}}=15$ and $ z=0.4$.}
\label{result6}
\end{figure}

In Fig. \ref{result6}, we show the   offloaded traffic $T$ of  our proposed  scheme  and the two baselines versus the content library size $F$ with ${\gamma ^{\text d}}=15$ and $ z=0.4$.
It can be observed that the   offloaded traffic of all the three schemes decreases with $F$.
The reason is that as $F$ becomes larger, the requested files will become more diverse and the number of requested files that are not cached locally will increase.
It can also be observed that the performance of the proposed scheme is superior to that of the baselines for all $F$.

\section{Conclusions}

In this paper, we have proposed a graph-based cooperative caching scheme including clustering and content placement in F-RAN.
By constructing the relevant node graph and  weighted graph,
the objective cluster sets  have been obtained by searching for the max-weight independent subset of the vertex set of the weighted graph.
By constructing the redundancy graph, the final caching decisions have been obtained by calculating the indicators of the caching-decision enhancements.
Both significant computational complexity reduction and  remarkable offloaded traffic have been achieved by using our proposed graph-based cooperative caching scheme.

\small
\bibliographystyle{IEEEtran}
\bibliography{manuscript-coopcaching}

\end{document}